\documentclass[preprint,showpacs,preprintnumbers,amsmath,amssymb]{revtex4}
\usepackage{graphicx}
\usepackage{dcolumn}
\usepackage{bm}
\usepackage{epsfig}

\begin{document}
\title{ The general behavior  of $NLO$  unintegrated parton distributions based on the single-scale evolution and the angular ordering constraint}
\author {H. Hosseinkhani}
\affiliation{Physics Department, University of Tehran, 1439955961
Tehran, Iran.}
\author{M. Modarres}\altaffiliation {  Email :
mmodares@ut.ac.ir, Tel:+98-21-61118645, Fax:+98-21-88004781}
 \affiliation{Physics Department, University of Tehran, 1439955961 Tehran, Iran.}

\begin{abstract}
To overcome the complexity of  generalized two hard scale
($k_t$,$\mu$) evolution equation, well known as the $Ciafaloni$,
$Catani$, $Fiorani$ and $Marchsini$ ($CCFM$) evolution equations,
and calculate the unintegrated parton distribution functions
($UPDF$), $Kimber$, $Martin$ and $Ryskin$ ($KMR$) proposed a
procedure based on ($i$) the inclusion of single-scale ($\mu$) only
at the last step of evolution and ($ii$) the angular ordering
constraint ($AOC$) on the $DGLAP$ terms (the $DGLAP$ collinear
approximation), to bring the second scale, $k_t$ into the $UPDF$
evolution equations. In this work we intend to use the $MSTW 2008$
(Martin et al) parton distribution functions (PDF) and try to
calculate $UPDF$ for various values of $x$ (the longitudinal
fraction of parton momentum), $\mu$ (the probe scale) and $k_t$ (the
parton transverse momentum) to see the general behavior of three
dimensional $UPDF$ at the  $NLO$ level up to the $LHC$ working
energy  scales ($\mu^2)$. It is shown that there exits some
pronounced peaks for the three dimensional $UPDF$ $(f_a(x,k_t))$
with respect to the  two variables $x$ and $k_t$ at various energies
($\mu$). These peaks get larger and move to larger values of $k_t$,
as the energy ($\mu$) is increased. We hope these peaks could be
detected in the $LHC$ experiments at $CERN$ and other laboratories
in the less exclusive processes.
\end{abstract}

 \maketitle

\section{INTRODUCTION}
To understand the event structure observed in different laboratories
i.e. $SLAC$, $HERA$, $DESY$ etc, and especially the one would be
expected in $LHC$ ($CERN$), the theoretical formalisms which
describe the small $x$ ($x$ is $Bejorken$ variable) region are
vital. The main unknown parameters  in these models are the
unintegrated parton distribution functions ($UPDF$)
\cite{3a,3b,3c,3d}. The $UPDF$ are two-scales dependent
distributions which are functions of $x$ (longitudinal momentum
fraction of the parent hadron) and the scales $k_{t}^2$ and
$\mu^{2}$, the squared transverse momentum of the parton and the
factorization scale, respectively. As we pointed out these
distributions are the  essential ingredients for the less exclusive
phenomenological computations in the high energy collisions of
particle physics.

It is well known that in the region of high energy and moderate
momentum transfer i.e. small $x$, the collinear factorization
theorem i.e.  $Dokshitzer$-$Gribov$-$Lipatov$-$Altarelli$-$Parisi$
($DGLAP$) \cite{5a,5b,5c,5d} evolution, breaks down. This happens
because of  the large increase of the phase space available for the
gluon emissions (i.e. a rapid rise in the gluon density), which
makes the quantum chromodynamics ($QCD$) perturbative expansions
unjustified and one can not obtain the $UPDF$. On the other hand, at
above high energy limit,  the cross section can be predicted by
using the $k_{t}$ factorization and the
$Balitsky$-$Fadin$-$Kuraev$-$Liptov$ ($BFKL$)
\cite{bfkl1,bfkl2,bfkl3} evolution. But the precision of $k_{t}$
factorization is not good e.g. the next-to-leading order ($NLO$)
corrections to $BFKL$ are very large \cite{bfkl4,bfkl5,bfkl6,bfkl7}.
Another approach to derive the $UPDF$ is the
$Ciafaloni$-$Catani$-$Fiorani$-$Marchesini$ ($CCFM$) equations
\cite{1a,1b,1c,1d,1e}. Although the $CCFM$ equations describe the
evolution of the $UPDF$ correctly, but working in this framework is
a complicated task, so practically they are used only in the Monte
Carlo event generators \cite{2b,2c,2d,2e,2f}. On the other hand, up
to now, there is not a complete quark version for these kind of
  equations \cite{1a,1b,1c,1d,1e,webber}, since   the
enhanced terms that are resumed by $CCFM$ come from gluon evolution.
However, to over come this problem, it has been shown that the
$CCFM$ equation can be reformulated (the linked dipole chain model)
by reducing the division between the initial and the final state
radiation diagrams using the colour dipole cascade model
\cite{LDCM1,LDCM2,Artem}.

The $Kimber$, $Martin$ and $Ryskin$ ($KMR$) \cite{4} approach is an
alternative prescription for producing the $UPDF$ which is based on
the standard $DGLAP$ equations \cite{5a,5b,5c,5d},
\begin{eqnarray}
\frac{\partial a(x,\mu^{2})}{\partial \ln(\mu^{2})}=\sum_{a'=q,g}
P_{aa'}\otimes a'(y,\mu^{2}),
 \label{1}
\end{eqnarray}
where $a(x,\mu^{2})=xq(x,\mu^{2})$ or $ xg(x,\mu^{2})$ and
$P_{aa'}(z)$ are the conventional (integrated) parton distribution
functions (PDF) and the well known $DGLAP$ splitting functions,
respectively. In equation (\ref{1}) the symbol $\otimes$ denotes a
convolution as,
\begin{eqnarray}
 f\otimes g=\int_x^{1}\frac{dy}{y}f(\frac{x}{y})g(y).
 \label{2}
\end{eqnarray}
In this approach under the certain approximation the $UPDF$ are
obtained from the PDF by introducing the scale $\mu$ only in the
last step of evolution with the inclusion of angular ordering
constraint ($AOC$). It has been shown that the $KMR$ prescription
gives the same results both for the $DGLAP$ and the unified
$BFKL$-$DGLAP$ equations \cite{4p} and the $AOC$ is applicable to
all orders as in the $CCFM$ formalism, i.e.
 all the loops contributions via the chain of evolution which are restricted by $AOC$, are resumed.

In this work, along  the lines of our recent calculations
\cite{8a,8b}, we intend to use the $KMR$ prescription with $MSTW
2008$ \cite{9} $PDF$ to produce three dimensional plots of $UPDF$ at
different energies ($\mu$) and discussed the various behavior of
$UPDF$ i.e. $f(x,k_t,\mu)$ . So the paper is organized as follows:
In section $II$ we briefly introduce the $KMR$ formalism and
finally, section $III$ is devoted to the results and the discussions
concerning the three dimensional (3D) graphs of the $UPDF$ produced
via this approach.
\section{The $KMR$ formalism \cite{4}}
The $KMR$ prescription \cite{4} works as a machine that by taking a
defined PDF as inputs, generates $UPDF$, as outputs. Using the
leading order (LO) splitting functions, $P_{aa'}$, the $DGLAP$
equations can be written in a modified form as \cite{4},
\begin{eqnarray}
\frac{\partial a( x,\mu^{2})}{\partial
\ln(\mu^{2})}=\frac{\alpha_s}{2\pi}\left[\int_x^{1-\Delta}P_{aa'}(z)\,a'\left(\frac{x}{z}
, \mu^{2}\right)dz-a(x ,
\mu^{2})\sum_{a'}\int_0^{1-\Delta}dz'P_{a'a}(z')\right],
 \label{3}
\end{eqnarray}
where $\Delta$ is a cutoff to prevent $z=1$ singularities in the
splitting functions arising from the soft gluon emission. In the
conventional $DGLAP$ formalism, $\Delta=0$ and the singularities are
canceled by the virtual terms. The value of $\Delta$ can be
determined by imposing an appropriate dynamical condition which is
replaced  by the angular ordering  constraint  arising  from the
coherency of the gluon emissions \cite{6a,6b},
\begin{eqnarray}
 ...>\theta_{n}>\theta_{n-1}>\theta_{n-2}> ...,
 \label{4}
\end{eqnarray}
where $\theta$'s are the radiation angels. This condition, at the
final step of  evolution, leads to \cite{1a,1b,1c,1d,4p},
\begin{eqnarray}
\mu>\frac{zk_t}{1-z}\Rightarrow
\Delta=1-z_{max}=\frac{k_t}{\mu+k_t}.
 \label{5}
\end{eqnarray}
The first part of the equation (\ref{3}), shows the contribution of
real emissions, that can change the transverse momentum $k_t$. The
second term expresses the evolutions due to the virtual effects
without changing the $k_t$. The latter can be re-summed, to obtain a
survival probability factor,
\begin{eqnarray}
T_a(k_t,\mu)=\exp\left[-\int_{k_t^2}^{\mu^2}\frac{\alpha_s({k'_t}^2)}{2\pi}
\frac{{dk'_t}^{2}}{{k'_t}^{2}}\sum_{a'}\int_0^{1-\Delta}dz'P_{a'a}(z')\right].
 \label{6}
\end{eqnarray}
Now, similar to the $Sudakov$ form factor, the above survival
probability, equation ({\ref{6}),  is imposed  into the equation
(\ref{1}), and by using equation (\ref{2}),  we find the equation
which describes the $UPDF$,
\begin{eqnarray}
f_{a}(x,k_{t}^2,\mu^{2})&=&T_a(k_t,\mu)\left[\left.
\frac{\partial\,a(x,\mu^{2})}{\partial\,ln(\mu^{2})}\right|_{\,\mu^2=k_t^2}\right]_{real}\nonumber\\
&=&T_a(k_t,\mu)\frac{\alpha_s({k_t}^2)}{2\pi}\int_x^{1-\Delta}P_{aa'}(z)\,a'\left(\frac{x}{z}
, {k_t}^2 \right)dz.
 \label{7}
\end{eqnarray}
More explicit forms of the above equation for the gluon $g$ and the
different quark flavors $q=u, d, s, ...$ are as follows,
\begin{eqnarray}
f_{q}(x,k_{t}^2,\mu^{2})&=&T_q(k_t,\mu)\frac{\alpha_s({k_t}^2)}{2\pi}
\nonumber\\&\times&
\int_x^{1-\Delta}dz\left[P_{qq}(z)\frac{x}{z}\,q\left(\frac{x}{z}
, {k_t}^2 \right) + P_{qg}(z)\frac{x}{z}\,g\left(\frac{x}{z} ,
{k_t}^2 \right)\right],
 \label{8}
\end{eqnarray}
and
\begin{eqnarray}
f_{g}(x,k_{t}^2,\mu^{2})&=&T_g(k_t,\mu)\frac{\alpha_s({k_t}^2)}{2\pi}
\nonumber\\&\times& \int_x^{1-\Delta}dz\left[\sum_q
P_{gq}(z)\frac{x}{z}\,q\left(\frac{x}{z} , {k_t}^2 \right) +
P_{gg}(z)\frac{x}{z}\,g\left(\frac{x}{z} , {k_t}^2 \right)\right].
 \label{9}
\end{eqnarray}
The key observation here is the dependency on the scale $\mu^2$,
which appears at the last step of the evolution. Another point is
that, the $Sudakov$ form factor which arises from the resumption of
virtual effects, can be used at every order of approximation.
Although the splitting functions must be used at the $NLO$ level,
but as it is shown in \cite{37},  the $NLO$ corrections to the
splitting functions, are relatively small in comparison to the $LO$
contributions. However, as stated above, only the $LO$ splitting
functions are used. On the other hand, although the definition of
$Sudakov$ form factor (like the $PDF$ themselves) has been started
intuitively from a probabilistic interpretation, but its role in the
mathematical description of the evolution remains in the equations.

The primary computations based on this kind of approach to evaluate
the $UPDF$, show very good agreement with the experimental data for
$F_2$ \cite{4}. Also, in recent years, the $KMR$ prescription have
been widely used for phenomenological calculations (see \cite{8a}
and the references therein). Recently the stability and the
reliability of the $KMR$ \ $UPDF$ have been investigated in
\cite{8a,8b}.

Finally, we should mention here that, the key property of the $CCFM$
approach (as given in their publications
\cite{1a,1b,1c,1d,1e,2b,2c}) is the $AOC$, which in turn has root in
the coherency of gluon radiation along the evolution chain, that is
valid for whole range of $x$ values. In the conventional $DGLAP$
formalism, the strong ordering constraint on the transverse momenta,
restricts the domain of study to the large and moderate values of
$x$:
$$
\hat{ \sigma}(\gamma^\ast q\rightarrow
qg)=\int_{p_{t_{min}}}^{p_{t_{max}}}dp^2_t{d\hat{\sigma}\over
{dp^2_t}},
$$
where
$$
p^2_{t_{min}}=\lambda^2,
$$
and
$$
p^2_{t_{max}}=p_t^2|_{\sin^2 \theta=1}=k^{\prime^2}={\hat{ s}\over
4}=Q^2{1-x\over 4x}.
$$
So to obtain the $DGLAP$ equations with $\ln({\hat{ s}\over
4})\simeq \ln (Q^2)$, $x$ should not be very low. In the $KMR$
prescription the $AOC$ property of the $CCFM$ formalism is applied
to modified $DGLAP$ evolution as a cut off on the integrals.
Therefore, the results of these modifications show that the effect
of application of $AOC$ is even more important than the inclusion of
the conventional low $x$ effects in the $BFKL$ approach \cite{4}.

\section{Results and discussion}
 As we stated in the section $II$, by using the equations (\ref{8}) and (\ref{9}), the $UPDF$ are
generated via the $KMR$ procedure. For the input $PDF$, the $MSTW
2008$ \cite{9} set of partons at the $NLO$ level are used
\footnote[1]{We use the $MSTW 2008$ code that is accessible from
\emph{http://durpdg.dur.ac.uk} and \emph{http://www.hepforge.org}.}.
Since the generated $UPDF$ ($f_{a}(x,k_{t}^2,\mu^{2})$) are three
variable functions, by fixing the scale $\mu^2$, their values versus
$x$ and $k_{t}^2$ are plotted in the various panels of figures 1, 2,
3 and 4 for the gluons, the up, the strange and the bottom  quarks,
respectively. For the better comparison, the values of the $\mu^2$
are chosen in a wide range $\mu^2=10, 10^{2}, 10^{4}, 10^{8}$
$GeV^2$ which is up to the $LHC$ working scales. The three typical
quark flavors, the $u$ quarks consists of the valence and the sea
contributions $u=u_v+u_{sea}$ and the $s$ and the $b$ quarks which
are completely sea distributions, are presented. The main feature of
these figures is exhibiting the general behavior of the $UPDF$ with
respect to the coupled contributions of $x$ and $k_t^2$. For
example, the most probable value of $k_t^2$ ($x$) at every $x$
($k_t^2$) for any kind of partons can be checked. As it can be seen,
by increasing the scale $\mu^2$ the graphs are shifted to the higher
$k_t^2$. This is expected, since the probability of finding partons
with larger $k_t^2$ is more probable at higher scales. The growth of
the values of the distributions by increasing $\mu^2$ and decreasing
$x$ and also the phenomenon of converging the quark distributions to
a unique value at small $x$ are known characteristics of the parton
distributions which are the heritage of their parent $PDF$. The
different behaviors of up and strange quarks at large $x$ have root
in the valence contribution in the case of up quark. The pronounced
peaks   become wider with respect to $k_t^2$, and move to higher
values of $k_t^2$. This behavior is much effective  for the up, the
strange and the bottom quarks. The peaks come from the concept of
distributions  and they are results of the dynamical evolution of
partons. The figures show that at given values of  hard scale and
$x$, at which $k_t$, it is more probable to detect the out going
partons. So based on the final partons, we can predict the dynamical
properties of the produced jets and their components, and on the
other hand it can inform us about the precision of the current
theoretical formalisms itself. The input PDF of $MSTW 2008$ are also
given in the figure 5, for comparison. With good approximation by
integrating over $UPDF$, we can get the input   $MSTW 2008$ PDF
($a(x,\mu^2)=\int^{\mu^2} {{dk_t^2\over k_t^2}}f_a(x,k_t^2,\mu^2)$).
For example for gluons, at $x=0.01$ and $\mu^2=100$ we get $6.7$
whereas $MSTW 2008$ gives the value of $6.5$ i.e. 3\% off.
Situations are the same for other points and parton distributions.
It is worth to say that in the original $KMR$ work, they get 25\%
discrepancies \cite{4} for above comparison. This is also evident by
comparison of figure 5 with those of 1 to 4 i.e. the $UPDF$ are
decreasing by increasing $x$. On the other hand, as have been
discussed in the $KMR$ and other related works, because of the
imposition of angular ordering, the $UPDF$ have values for
$k_t^2\geq \mu^2$ as $x$ decreases. But this will not affect the
above integration too much. The figures 1 to 4 also show that, for
low scales ($\mu^2\simeq 10$ $GeV^2$) the $UPDF$ become negative
when $x$ becomes close to one. This reflects the negative values of
$MSTW 2008$ gluon distributions at the $NLO$ level and beyond that.
So the negative values of $UPDF$ have root in the parent integrated
gluon distributions which in turn are the result of $MSTW 2008$
assumptions \cite{9}.  As it was pointed, in the $MSTW2008$
\cite{9}, for better data fitting it is allowed that, the gluon
distribution takes negative values, because there is no theorem that
imposes positivity condition on $PDF$ beyond the  $LO$
approximation. So they become negative  in order to fit the data (in
other words they can be traced to the slow evolution of $F_2$ at
small $x$ and  $Q^2$ i.e. a positive gluon would give too rapid
evolution to fit the $dF_2/d\ln(Q^2)$ data. Then in the $KMR$
integrals, the evaluation of input $g(x,k_t^2)$ at small $x$ and
$k_t$ (as a scale, instead of $Q^2$  in $g(x,Q^2)$) leads to the
negative values for the output $UPDF$. Finally, $(i)$ the comparison
of $UPDF$ produced from different  $PDF$ sets have been made in our
former works \cite{8a,8b} . The different parameterizations
procedures lead to different $PDF$, and a discussion about these
procedures is presented in \cite{8a,8b} and references therein.
$(ii)$ The differences between  the $LO$ and the $NLO$ $PDF$ are
parameterizations dependent. In the $MSTW2008$ this is noticeable,
but in some other parameterizations sets based on different
assumptions and procedures it can be less (e.g $GRV$ sets
\cite{8a,8b}), but as we have showed in \cite{8a,8b} (by
investigating the ratios of $KMR$ $UPDF$ compared to the
corresponding ratios of input $PDF$) the relative differences are
less in the output $UPDF$ and the $KMR$ prescription suppresses
these discrepancies. To show this point more transparently,  in
figure 6 we have plotted the gluon $UPDF$ with three different input
$PDF$, namely the original $KMR$ \cite{4} with $MRST99$ \cite{99}
$PDF$, our recent works \cite{8a,8b} with $GJR08$ $PDF$ \cite{GRV}
and present calculation ($MRST2008$) at $\mu^2=100$ $GeV^2$ and
$x=0.1, 0.01, 0.001$ and $0.0001$ in terms of $k_t^2$. It is clearly
seen that different input $PDF$ give very similar $UPDF$. $(iii)$ In
fact a complete prescription for producing the $NLO$ $UPDF$ needs to
include both the $PDF$ and the splitting functions at the $NLO$
level. This prescription is presented in \cite{37}, but as it is
shown in this reference \cite{37}, inclusion of the $NLO$ splitting
functions have very low effect comparing to the contribution of the
$NLO$ $PDF$. Therefore, ignoring the corrections due to the $NLO$
splitting functions do not affect our analysis of the general
behavior of the $NLO$ $UPDF$. $(iv)$ There is no restriction on the
$k_t$ dependency. As the orders of the approximation are in terms of
orders of $\alpha_s(k_t^2)$, the $NLO$ accuracy is contained in the
NLO PDF and splitting functions that discussed in the former
comments. Hence at scales $k_t^2\geq Q_0^2$ ,where $Q_0^2$ is the
scale that upper than it, the perturbative $QCD$ is still
applicable, these results are valid.

\begin{acknowledgments}
We would like to acknowledge  the Research Council of University of
Tehran and Institute for Research and Planning in Higher Education
for the grants provided for us. $HH$ would like to thank Professors
$Hannes$ $Jung$, $Francesco$ $Hautmann$, $Artem$ $Lipatov$, $Gavin$
$Salam$, $James$ $Stirling$, $Graeme$ $Watt$ and $Bryan$ $Webber$
for their helpful communications.
\end{acknowledgments}

\newpage
\begin{figure}[ht]
\caption{The unintegrated gluon distribution functions generated by
the $KMR$ procedure with the fixed values of $\mu^2=10, 10^{2},
10^{4}, 10^{8}$ $GeV^2$.}
\end{figure}

\begin{figure}[ht]%
\caption{As figure 1 but for the up quark.}
\end{figure}

\begin{figure}[ht]%
\caption{As figure 1 but for the strange quark}
\end{figure}

\begin{figure}[ht]%
\caption{As figure 1 but for the bottom quark}
\end{figure}

\begin{figure}[ht]%
\caption {The $NLO$ integrated parton distribution function of $MSTW
2008$ versus $x$ for the fixed values of $\mu^2=10, 10^{2}, 10^{4},
10^{8}$ $GeV^2$}
\end{figure}

\begin{figure}[ht]%
\caption {The $UPDF$ of $MSTW 2008$ (present calculation, dotted
curve), $MRST99$ (dash curve) and $GJR08$ (full curve). See the text
for details. }
\end{figure}


\begin{thebibliography}{35}

\bibitem{3a} B. Andersson et al. (Small-x Collaboration),
Eur.Phys.J.C, {\bf 25} (2002) 77.

\bibitem{3b} J. Andersen et al. (Small-x Collaboration), Eur.Phys.J.C, {\bf 35} (2004) 67.

\bibitem{3c} J. Andersen et al. (Small-x Collaboration), Eur.Phys.J.C, {\bf 48} (2006) 53.

\bibitem{3d} H. Jung et al., Proceedings of HERA and the LHC workshop on the implications
of HERA for LHC physics, 2006 - 2008, Hamburg - Geneva ,(2009);
arXiv:0903.3861.

\bibitem{5a} V.N. Gribov and L.N. Lipatov, Yad.Fiz., {\bf 15}
(1972) 781.

\bibitem{5b} L.N. Lipatov, Sov.J.Nucl.Phys., {\bf  20} (1975) 94.

\bibitem{5c} G. Altarelli and G. Parisi, Nucl.Phys.B, {\bf 126} (1977) 298.

\bibitem{5d} Y.L. Dokshitzer, Sov.Phys.JETP, {\bf 46} (1977) 641.

\bibitem{bfkl1} E.A. Kuraev, L.N. Lipatov, and V.S. Fadin, Sov.Phys.JETP, {\bf
44} (1976) 443.

\bibitem{bfkl2} E.A. Kuraev, L.N. Lipatov and  V.S. Fadin, Sov.Phys.JETP, {\bf 45} (1977) 199.

\bibitem{bfkl3} I.I. Balitsky andL.N. Lipatov, Sov.J.Nucl.Phys., {\bf  28} (1978) 822.

\bibitem{bfkl4}L.V. Gribov, E.M. Levin and M.G. Ryskin, Phys.Rept., {\bf 100} (1983) 1.

\bibitem{bfkl5} E.M. Levin, M.G. Ryskin, Y.M. Shabelski and A.G. Shuvaev, Sov.J.Nucl.Phys., {\bf  53} (1991) 657.

\bibitem{bfkl6} S. Catani, M. Ciafaloni and F. Hautmann, Nucl.Phys.B, {\bf 366} (1991) 135.

\bibitem{bfkl7}J.C. Collins and  R.K. Ellis, Nucl.Phys.B, {\bf 360} (1991) 3.

\bibitem{1a} M. Ciafaloni, Nucl.Phys.B, {\bf 296} (1988) 49.

\bibitem{1b} S. Catani, F. Fiorani, and G. Marchesini, Phys.Lett.B, {\bf 234} (1990) 339.

\bibitem{1c} S. Catani, F. Fiorani, and G. Marchesini, Nucl.Phys.B {\bf 336} (1990) 18.

\bibitem{1d} G. Marchesini, in Proceedings of the Workshop
   ''QCD at 200 TeV,'' Erice, Italy, edited by L. Cifarelli and
   Yu.L. Dokshitzer, Plenum, New York, (1992) 183.

\bibitem{1e} G. Marchesini, Nucl.Phys.B, {\bf 445} (1995) 49.

\bibitem{2b} G. Marchesini and B. Webber, Nucl.Phys.B, {\bf 349} (1991) 617.

\bibitem{2c} G. Marchesini and B. Webber, Nucl.Phys.B, {\bf 386} (1992) 215.

\bibitem{2d} H. Jung, Nucl.Phys.B, {\bf  79} (1999) 429.

\bibitem{2e} H. Jung and G.P. Salam, Eur.Phys.J.C, {\bf 19} (2001) 351.

\bibitem{2f} H Jung, J.Phys.G:Nucl.Part.Phys., {\bf 28} (2002) 971.

\bibitem{webber} B.  Webber, private communication (2010).

\bibitem{LDCM1} H. Kharraziha and L. L\"{o}nnblad, JHEP {\bf 03}(1998) 006.

\bibitem{LDCM2}B. Andersson, G. Gustafson and J. Samuelsson, Nucl.Phys.B, {\bf 463} (1996) 215.

\bibitem{Artem}Artem Lipatov and Gavin Salam private communications (2010).

\bibitem{4} M.A. Kimber, A.D. Martin and M.G. Ryskin, Phys.Rev.D, {\bf 63} (2001)
114027.

\bibitem{4p} M.A. Kimber, J. Kwiecinski A.D. Martin and A.M. Stasto, Phys.Rev.D, {\bf 62} (2000) 094006.

\bibitem{9} A.D. Martin, W.J. Stirling, R.S. Thorne and G. Watt, Eur.Phys.J.C, {\bf 63} (2009) 189.

\bibitem{8a} M. Modarres and H. Hosseinkhani , Few-Body Syst., {\bf 47} (2010) 237.

\bibitem{8b} M. Modarres and H. Hosseinkhani, Nucl.Phys.A, {\bf 815} (2009) 40.


\bibitem{6a} G. Marchesini and B.R. Webber, Nucl.Phys.B, {\bf 310} (1988) 461.

\bibitem{6b} Yu.L. Dokshitzer, V.A. Khoze, S.I. Troyan and A.H. Mueller, Rev.Mod.Phys., {\bf 60} (1988) 373.

\bibitem{37} A.D. Martin, M.G. Ryskin and  G. Watt, Eur.Phys.J.C,
{\bf 66} (2010) 163.

\bibitem{99} A.D. Martin, W.J. Stirling, R.S. Thorne and G. Watt, Eur.Phys.J.C, {\bf 14} (2000) 133.

\bibitem{GRV}M.  Gl\"{u}ck, P. Jimenez-Delgado and E. Reya, Eur.Phys.J.C {\bf 53}
(2008) 355.

\end{thebibliography}
\end{document}